\documentclass[11pt,a4paper]{article}
\usepackage[latin1]{inputenc}
\usepackage{amsmath}
\usepackage{amsfonts}
\usepackage{amssymb}
\usepackage{graphicx}
\usepackage{caption}
\usepackage{subcaption}
\usepackage{comment}
\usepackage[normalem]{ulem}
\usepackage{array}
\usepackage{makecell}
\usepackage{xspace}
\usepackage{soul}
\usepackage[left=2.00cm, right=2.00cm, top=2.00cm, bottom=3.00cm]{geometry}

\newcommand*{\bquark}{\ensuremath{b\text{-quark}}\xspace}
\newcommand*{\bquarks}{\ensuremath{b\text{-quarks}}\xspace}
\newcommand*{\GeV}{\ensuremath{\text{Ge\kern -0.1em V}}}
\newcommand*{\tev}{\ensuremath{\text{Te\kern -0.1em V}}}
\newcommand*{\TeV}{\ensuremath{\text{Te\kern -0.1em V}}}
\newcommand*{\antibar}[1]{\ensuremath{#1\bar{#1}}\xspace}
\newcommand*{\bbbar}{\antibar{b}}
\newcommand*{\ttbar}{\antibar{t}}
\newcommand*{\MET}{\ensuremath{E_{\text{T}}^{\text{miss}}}\xspace}
\newcommand*{\DeltaR}{\ensuremath{\Delta R}\xspace}
\newcommand*{\pt}{\ensuremath{p_{\text{T}}}\xspace}
\newcommand*{\ifb}{\mbox{fb\(^{-1}\)}}

\newcommand{\cHWtil}{\tilde{c_{HW}}}
\newcommand{\oHWtil}{\tilde{O_{HW}}}

\newcommand{\ptw}{ p_T^W }

\newcommand{\mttot}{ m_{T}^{\ell \nu b \bar{b}} }
\newcommand{\pznu}{p_z^\nu}
\newcommand{\ptnu}{p_T^\nu}
\newcommand{\cDPlus}{\cos \delta^+}
\newcommand{\cDMinus}{\cos \delta^-}
\newcommand{\qlCDPlus}{Q_\ell \cDPlus}
\newcommand{\qlCDMinus}{Q_\ell \cDMinus}

\usepackage[dvipsnames]{xcolor}
\usepackage{hyperref}
\hypersetup{
	colorlinks=true,
	citecolor=blue
}

\usepackage{tcolorbox}

\graphicspath{{Figures/}}

\title{Simulation-based inference in the search for CP violation\\ in leptonic WH production}

\date{\today}
\author{
Ricardo Barru\'e$^{1, 2\,}$\footnote{E-mail: \texttt{ricardo.barrue@cern.ch}},
Patricia Conde Mu\'{\i}\~no$^{1, 2\,}$\footnote{E-mail:
	\texttt{patricia.conde.muino@cern.ch}},
Valerio Dao$^{3\,}$\footnote{E-mail:
	\texttt{valerio.dao@cern.ch}},
Rui Santos$^{4,5\,}$\footnote{E-mail:  \texttt{rasantos@fc.ul.pt}}
\\[9mm]
{\small\it
$^1$Laborat\'orio de Instrumenta\c c\~ao e F\'{\i}sica Experimental de Part\'{\i}ulas - LIP,}\\
{\small\it Av. Gama Pinto, 2, 1649-003 Lisboa, Portugal.} \\[3mm]
{\small\it
$^2$Departamento de F\'{\i}sica, Instituto Superior T\'ecnico, Universidade de Lisboa,}\\
{\small\it Av. Rovisco Pais, 1049-001, Lisboa, Portugal.} \\[3mm]
{\small\it
$^3$Department of Physics and Astronomy, Stony Brook University;}\\
{\small\it NY, United States of America.} \\[3mm]
{\small\it
$^4$ISEL -
 Instituto Superior de Engenharia de Lisboa,} \\
{\small \it   Instituto Polit\'ecnico de Lisboa
 1959-007 Lisboa, Portugal.} \\[3mm]
{\small\it
$^5$Centro de F\'{\i}sica Te\'{o}rica e Computacional,
    Faculdade de Ci\^{e}ncias,} \\
{\small \it    Universidade de Lisboa, Campo Grande, Edif\'{\i}cio C8
  1749-016 Lisboa, Portugal.} \\[3mm]
}

\begin{document}
	\maketitle
	
	\begin{abstract}
Sources of CP violation beyond the Standard Model (BSM) are required to explain the baryonic asymmetry of the Universe. In this work, we study BSM CP-violating components in the HWW interaction in WH production, parametrized by an effective dimension-6 CP-odd operator. We explore a machine learning simulation-based inference method that estimates a detector-level optimal observable - SALLY - comparing it with energy-dependent and angular observables, exploring different binnings for their distributions. We show that in regions of phase space where the interference between SM and the effective operator dominates, a SALLY observable leads to optimal limits. In regions where effects of the quadratic term of the effective operator start becoming dominant, such an observable still leads to optimal limits. This work aims to test current multivariate techniques and inform analysis strategies for LHC Run 3 and beyond.

\end{abstract}

\thispagestyle{empty}
\vfill
\newpage
\setcounter{page}{1}
	
	\section{Introduction}\label{sec:intro}

Violation of charge-parity (CP) symmetry from beyond the Standard Model (BSM) physics is required to explain the observed matter-antimatter asymmetry of the Universe. Higgs boson interactions are a natural place to search for BSM CP-violating components, given that the effect of such components on inclusive cross-sections is negligible when compared with that coming from BSM CP-conserving components. The same happens to their impact on cross-sections differential in energy-related observables, such as the ones targetted by the current Simplified Template Cross Section formalism \cite{VH_resolved_run2_ATLAS}. Searches for BSM CP-violating components typically use one of two approaches: either measuring differential cross-sections in bins of angular observables \cite{htautau_CP_ATLAS,ATLAS_VBF_Hgamgam_CP_diffXS,tthbb_CP_ATLAS}, or in bins of (statistically) optimal observables built from matrix element calculators and the complete kinematic information of the event \cite{VBFHTauTauOptObs2020ATLAS,MELAH4l2020CMS,ATLAS_VBF_Hgamgam_CP_optobs,CMS_VBF_Htautau_CP}. Both have sub-optimal points: the former measures only one or two observables simultaneously, discarding a significant fraction of the available information. The latter commonly requires neglecting or approximating parton shower, hadronization, and detector effects, often calculating matrix elements directly from detector-level observables. Refs. \cite{MadMinerIntro,MadMinerPhysicsComplete,MadMinerStat} introduce a machine-learning (ML) method which to estimate detector-level optimal observables (known as \textit{score} in statistics literature), called SALLY (Score Approximates Likelihood LocallY), which removes the need for such approximations.

In this work, we studied CP-violating components in the interaction between the Higgs boson and W boson pairs (HWW interaction) in the WH production channel, with the W boson decaying to leptons and the Higgs decaying to a pair of \bquarks. This channel is challenging because the neutrino longitudinal momentum and the W boson 4-vector cannot be reconstructed unambiguously. Nonetheless, they can be estimated, typically with significant uncertainties, that propagate to angular or matrix-element-based optimal observables.

We perform an analysis to compare the sensitivity of a SALLY observable with that of {energy-dependent and angular} observables, disentangling their sensitivity to different terms in the squared matrix element. We also study different binnings for all the observables.

The structure of this paper is as follows: Sec. \ref{sec:score_sbi} explains the theory behind the SALLY method. Sec. \ref{sec:EFT} introduces the effective field theory formalism, used in this analysis to parametrize CP-violating components and describes the current best measurements from experiments. Sec. \ref{sec:analysis} presents the analysis details: signal and relevant backgrounds, sample generation, and selection criteria. Sec. \ref{sec:obs} describes the different observables to be compared. Sec. \ref{sec:stat_analysis} explains the statistical methods used to extract the sensitivity of different observables. In Sec. \ref{sec:results}, we compare the sensitivity of the different observables and derive conclusions. 

\section{The SALLY method}\label{sec:score_sbi}

The (per-event) kinematic likelihood $p(x|\theta)$ is the central object of any particle physics analysis, where $x$ and $\theta$ are the reconstructed detector-level observables and the parameters of interest, respectively. The score, defined as $t(x|\theta_{ref}) = \nabla_\theta \log p(x|\theta)\rvert_{\theta_{ref}}$ is, by construction, the optimal observable for values of $\theta$ close to a reference value, $\theta_{ref}$. For a high-dimensional $x$, estimating the likelihood and the score is computationally challenging.  

The SALLY method starts from the joint likelihood $p(x,z|\theta)$ - \textit{joint} function of $x$ and the internal (latent) variables of the generation process, $z$. It calculates for each event the joint score, $t(x,z|\theta) = \nabla_\theta \log p(x,z|\theta)$, as seen in Eq. \ref{eq:jointScore}.

\begin{equation} \label{eq:jointScore}
	t(x,z|\theta_{ref}) = \frac{\nabla_\theta p(z_p|\theta)}{p(z_p|\theta)}\rvert_{\theta_{ref}} \approx \left[\frac{\nabla_\theta d\sigma(z_p|\theta)}{d\sigma(z_p|\theta)} - \frac{\nabla_\theta \sigma(\theta)}{\sigma(\theta)}\right]\rvert_{\theta_{ref}},
\end{equation}

\noindent where $z_p$ and $p(z_p|\theta) = d\sigma(z_p|\theta)/\sigma(\theta)$ are, respectively, the parton-level kinematics and parton-level likelihood, $d\sigma(z_p|\theta)$ and $\sigma(\theta)$ are, respectively, the event weights (proportional to the amplitude $|\mathcal{M}(z_p|\theta)|^2$) and cross-sections for a given $\theta$, extracted from the generator for each event. The SALLY method then uses the joint score in the loss function of a neural network with detector-level observables as input variables. Such an estimator will converge to the detector-level optimal observable in the limit of infinite training data. The joint score is equivalent to the standard Optimal Observable when neglecting the detector response and setting $\theta_{ref}=0$. SALLY is one of several \textit{simulation-based inference} methods, which use information from event generators to build estimators of the likelihood, likelihood ratio, or score.

There are other ML-based methods to reconstruct detector-level optimal observables for high-dimensional $x$ without approximations \cite{CP_asymmetries_Higgs_Pilkington,CP_asymmetries_EW_Pilkington,JHUGEN_paper}. These build an observable from the output of classifiers trained to distinguish between events with the same (fixed) value of the Wilson coefficient but with opposite sign. They differ from SALLY since they don't estimate the score and don't use generator information in the loss function.

\section{Effective field theory}\label{sec:EFT}

So far, searches at the LHC have not produced any compelling hints of the existence of new particles. This fact has boosted the use of the Effective Field Theory (EFT) approach to look for new physics. 
In the EFT formalism, the SM Lagrangian is extended with additional operators of mass dimension $d > 4$.  CP-violating effects come from the interference between terms in the SM Lagrangian and CP-odd higher dimension operators. These terms do not change the inclusive cross-sections, but lead to modifications in observable distributions. 

This work uses the Standard Model Effective Field Theory (SMEFT) \cite{SMEFT} formalism. In the SMEFT, the $d>4$ operators are built from combinations of SM fields, constrained to be invariant under the SM $SU(3)_c \times SU(2)_L \times U(1)_Y$ symmetry. Electroweak symmetry breaking happens as in the SM (linearly), where the massless $W$ and $B$ fields absorb a Goldstone boson to gain their longitudinal polarization/mass, leaving a physical state - the SM Higgs boson. The non-redundant operator basis defined in Ref. \cite{WarsawBasis} is used (the so-called Warsaw basis). The only dimension-6 CP-odd operator in the HWW vertex is $\tilde{\mathcal{O}}_{HW}$, defined in Eq. \ref{eq:OHWtil}.

\begin{equation}\label{eq:OHWtil}
	\tilde{\mathcal{O}}_{HW} = \frac{\cHWtil}{\Lambda^2} H^\dagger H \tilde{W}^I_{\mu\nu} W^{I\mu\nu}  = \frac{\cHWtil}{\Lambda^2} H^\dagger H \epsilon_{\mu\nu\rho\sigma} W^{I\rho\sigma} W^{I\mu\nu}\, .
\end{equation}


The best limits on $\cHWtil$ from the ATLAS and CMS collaborations are described in Refs. \cite{ATLAS_VBF_Hgamgam_CP_optobs} and \cite{CMS_VBF_Htautau_CP}, respectively.
In Ref. \cite{ATLAS_VBF_Hgamgam_CP_optobs}, the Optimal Observable \cite{OO1,OO2,OO3} technique is applied to VBF production in the $H \to \gamma \gamma$ decay, the observed 95\% CL limit is set at $\cHWtil \in [-0.53, 1.02]$ when only the SM-EFT interference component is taken into account and $\cHWtil \in [-0.55, 1.07]$ when quadratic EFT components are also taken into account. The bounds assume  $\Lambda= 1\:\TeV$.
In Ref. \cite{CMS_VBF_Htautau_CP}, the Matrix Element Likelihood Analysis (MELA) \cite{MELA1,MELA2,MELA3,MELA4,JHUGEN_paper} approach is used, with discriminants built from probably density functions extracted from matrix element calculators. In this analysis, the limits are placed on the effective CP-odd cross-section fraction (assuming $SU(2)_L \times U(1)_Y$ and custodial symmetries) at $f_{a3} \in [-0.01, 1.28] \times 10^{-3}$ at 95\% CL, translated to $\cHWtil \in [-1.16,0.1]$ using JHUGenLexicon \cite{JHUGEN_paper}, and setting $\Lambda= 1\:\TeV$.
	
	\section{Analysis}\label{sec:analysis}

The \texttt{MadMiner} package \cite{MadMiner_code} wraps around the entire analysis workflow and was used to develop this work (version 0.9.3).

\subsection{Signal and backgrounds}\label{sec:signals_bkgs}

\begin{figure}[h!]
	\centering
	\includegraphics[width=0.4\linewidth]{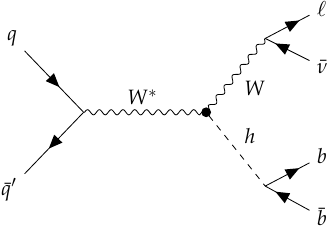}
	\caption{Feynman diagram for WH associated production in the $\ell \nu \bbbar$ final state. The vertex of interest is circled in black.}
	\label{fig:whlvbb}
\end{figure}

This analysis targets the associated WH production channel in the $W \to \ell \nu, H \to \bbbar$ final state $(\ell = e,\mu)$, represented by the Feynman diagram in Fig. \ref{fig:whlvbb}. The main reasons to choose this particular final state are the following: the decay of the Higgs boson to \bquarks is the one with the highest branching ratio $(BR \approx 58\%)$, and there is a high energy (isolated) lepton from the W decay which allows for efficient triggering and removal of a large part of the QCD multijet background. 

The HWW interaction can also be probed in VBF production and in the $H \to WW$ decay. Despite the higher cross-section of VBF production, it is a process where it is not possible to disentangle the contributions of the HZZ and HWW interaction vertices. The WH production channel allows access to the HWW vertex independently of the HZZ vertex, which removes the need for the assumption that custodial symmetry is not broken.

\begin{figure}[h!]
	\centering
	\begin{subfigure}[b]{0.28\textwidth}
		\centering
		\includegraphics[width=\textwidth]{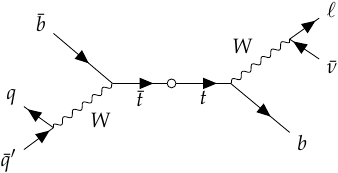}
		\caption{}
	\end{subfigure}
	\hfill
	\begin{subfigure}[b]{0.28\textwidth}
		\centering
		\includegraphics[width=\textwidth]{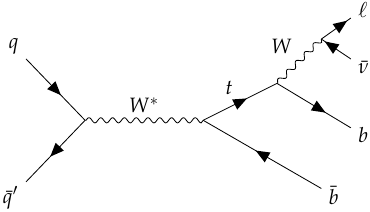}
		\caption{}
	\end{subfigure}
	\hfill
	\begin{subfigure}[b]{0.14\textwidth}
		\centering
		\includegraphics[width=\textwidth]{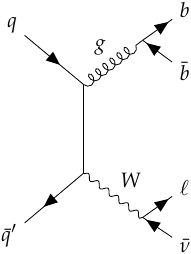}
		\caption{}
	\end{subfigure}
	\caption{Feynman diagrams for the main backgrounds in WH associated production: semileptonic $\ttbar$ (left), single top production in the $s$-channel (middle) and associated production of a W boson and $b$-jets (right).}
	\label{fig:backgrounds}
\end{figure}

The main backgrounds taken into account are $\ttbar$ production in the semileptonic decay channel, single top production in the $s$-channel, and associated production of a W boson and $b$-jets since these have the same or similar final state as the signal process. Their Feynman diagrams are shown in Fig. \ref{fig:backgrounds}.	

\subsection{Sample generation and selection conditions}\label{sec:samples}

We generated signal samples at LO in QCD with \texttt{Madgraph5\_aMC@NLO}~\cite{Madgraph5_aMC@NLO} using the \texttt{SMEFTsim3}~\cite{SMEFTsim3} UFO model assuming a $U(3)^5$ flavor symmetry. The experimental values of  
$m_W = 80.387$ GeV, $m_Z= 91.1876$ GeV and $G_F = 1.1663787 \times 10^{-5}$, $m_t = 172.76$ GeV,  $m_b = 4.18$ GeV, were used as input parameters, and other fermions were considered massless. We didn't perform any truncation of the squared matrix element and fixed the decay widths of all the relevant massive particles to their LO values. We set the new physics scale to $\Lambda = 1 \:\TeV$
We generated a signal sample with $30 \times 10^6$ events at $\cHWtil=0$. We used LO reweighting~\cite{Madgraph_reweighting} to calculate event weights for two values of $\cHWtil \neq 0$, and a morphing technique~\cite{morphing_paper_2022} to obtain the weights for all other values. We used the functionality in the \texttt{MadMiner} package to determine the values of $\cHWtil \neq 0$ which minimize the sum of the squared morphing weights, to avoid numerical instabilities from the morphing procedure. We set the maximum possible range to $|\cHWtil| \leq 1.2$ (slightly looser than the combination of experimental constraints shown in Sec. \ref{sec:EFT}), and the optimal points were set at $\cHWtil = 1.15$ and $\cHWtil = -1.035$. We used the PDF4LHC15 PDF set \cite{PDF4LHC15}.

For each of the backgrounds, we generated samples with $10 \times 10^6$ events at LO in QCD with the default \texttt{Madgraph5\_aMC@NLO} SM UFO model. We did not apply reweighting or morphing to the background samples, given that $\oHWtil$ does not affect these processes.

For parton shower and hadronization, we used Pythia with the A19 tune and MLM merging, setting the merging scale (that separates hard and parton shower emissions) to $Q_{cut} = 20 \:\GeV$.

After generating the samples, we passed the events through detector simulation, physics object identification, and selection using Delphes \cite{Delphes} with the default ATLAS card. We selected events where all the objects were reconstructed and the two leading jets were b-tagged, using the b-tagging efficiency map in the ATLAS Delphes card. These criteria retained $\approx 24\%$ of the generated events.

We applied the set of generator-level cuts first introduced in Ref. \cite{MadMinerWH} to mimic typical experimental analysis selection conditions. Table \ref{tab:selection_cuts} describes the cuts and their values.

\begin{table}[h!]
	\setlength{\extrarowheight}{1mm}

	\begin{tabular}{|l|r|}

		\textbf{Observable} & \textbf{Cut} \\ \hline
			Transverse momentum of lepton/light quarks (charm or lighter) & $p_{T,\ell}, p_{T,j} > 10 \:\GeV$ \\
			Missing transverse energy & $\MET > 25 \:\GeV$ \\
			Transverse momentum of \bquarks & $p_{T,b} > 35 \:\GeV$ \\
			Pseudorapidity of charged lepton, \bquarks and light quarks & $|\eta_{\ell,b,j}| < 2.5$ \\
			Angular distance between decay particles & $\DeltaR_{b b,b \ell,bj,\ell j, jj} > 0.4$ \\
			Invariant mass of \bquark pair & $80 \:\GeV < m_{bb} < 160 \:\GeV$ \\
			Transverse momenta of light quarks & $p_{T,j} < 30 \:\GeV$ \\ 
	\end{tabular}
\caption{Analysis selection conditions and the cut values, applied at generator-level.}
\label{tab:selection_cuts}
\end{table}

Before any cut is applied, the cross-sections are $213.6$ fb for the WH signal, $6.9 \times 10^4 $ fb for the $t\bar{t}$ background, $5.2 \times 10^4 $ fb for the W+(b-)jets background and $665.8 $ fb for the single top background. Table \ref{tab:cutflow} presents the cumulative efficiency (in \%) of the cuts up to a specific cut. After all of the cuts, the cross-sections are $73.16$ fb for the WH signal, $194.24$ fb for the $t\bar{t}$ background, $239.5$ fb for the W+(b-)jets background and $75.74$ fb for the single top background. The criteria with the largest background rejection are the cut on the minimum transverse momentum of the b-quark for W+jets and the one on the maximum value for the transverse momentum of light jets for $t\bar{t}$, with individual (background) rejection factors of 29 and 48, respectively. Given that the NLO corrections for effective Lagrangians with CP-odd operators are still unknown, we only used LO values for the cross-sections and did not employ any signal or background k-factors.

\begin{table}[h]
	\centering
	\setlength{\extrarowheight}{1mm}
	\begin{tabular}{l|r|r|r|r}
	Cut                             & Signal (SM) & $t\bar{t}$ & $W$+jets & Single top \\ \hline
	$p_{T_\ell}, p_{T_j} > 10 \:\GeV$                 & 96.77               & 87.12 & 93.83  & 93.83      \\
	$E_{T_\text{miss}} > 25 \:\GeV$              & 76.17               & 70.03 & 56.17  & 74.41      \\
	$p_{T_b} > 35 \:\GeV$                 & 50.05               & 52.08 & 1.91   & 50.6       \\
	$|\eta_{\ell,b,j}|  <   2.5$       & 35.42               & 39.14 & 1.25   & 35.01      \\
	$\Delta R_{bb,b \ell,bj,\ell j,jj} > 0.4$                      & 34.18               & 36.46 & 0.99   & 33.9       \\
	$80 \:\GeV < m_{bb} < 160 \:\GeV$            & 34.31               & 13.2  & 0.46   & 11.39      \\
	$p_{T_j} < 30 \:\GeV$                  & 34.25               & 0.28  & 0.46   & 11.38      \\ \hline
	\end{tabular}
	\caption{Cumulative efficiencies (in \%) of the cuts up to the specific cut in each line.}
	\label{tab:cutflow}
\end{table}
	
	\section{Observables}\label{sec:obs}

\subsection{Energy-dependent observables}\label{sec:energy_obs}

\begin{figure}[h!]
	\centering
	\includegraphics[width=0.75\linewidth]{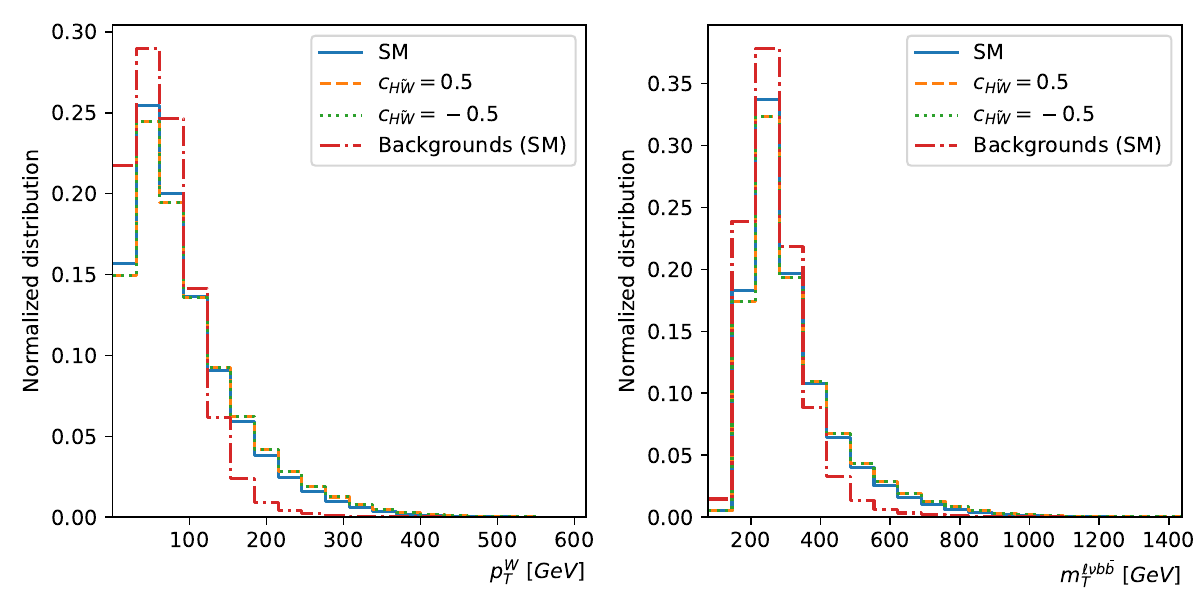}
	\caption{Distributions (normalized to unit area) of $\ptw$ (left) and $\mttot$ (right) for the SM backgrounds(red), SM signal (blue), and two signal samples with $\cHWtil=0.5$ (orange) and $\cHWtil=-0.5$ (green) obtained by morphing the benchmark signal samples.}
	\label{fig:ptwmttotshapeonly}
\end{figure}

The (CP-violating) SM-EFT interference component in the squared matrix element has a residual dependence on the partonic center-of-mass energy \cite{GodboleHCP,MethodOfMoments,wh_lvaa_eft_fcc}. Hence, it makes sense to start from observables that capture this dependence. A natural candidate is the transverse momentum of the $W$ boson, $\ptw$. Ref.~\cite{MadMinerWH} showed that the total transverse mass of the event, $\mttot$, captures energy-dependent effects complementary to $\ptw$. We show the distribution of these observables at the reconstruction level (normalized to unit area) in Fig.~\ref{fig:ptwmttotshapeonly}.
Both observables are sensitive to $\cHWtil \neq 0$, as demonstrated by the increase in the signal-to-background ratio in the high energy region compared to the SM prediction. Nonetheless, these are not sensitive to the sign of $\cHWtil$, which leads one to conclude that the modifications to these observables come mainly from the quadratic EFT term in the squared matrix element.

\subsection{Angular observables}\label{sec:ang_obs}

\begin{figure}
	\centering
	\includegraphics[width=\linewidth]{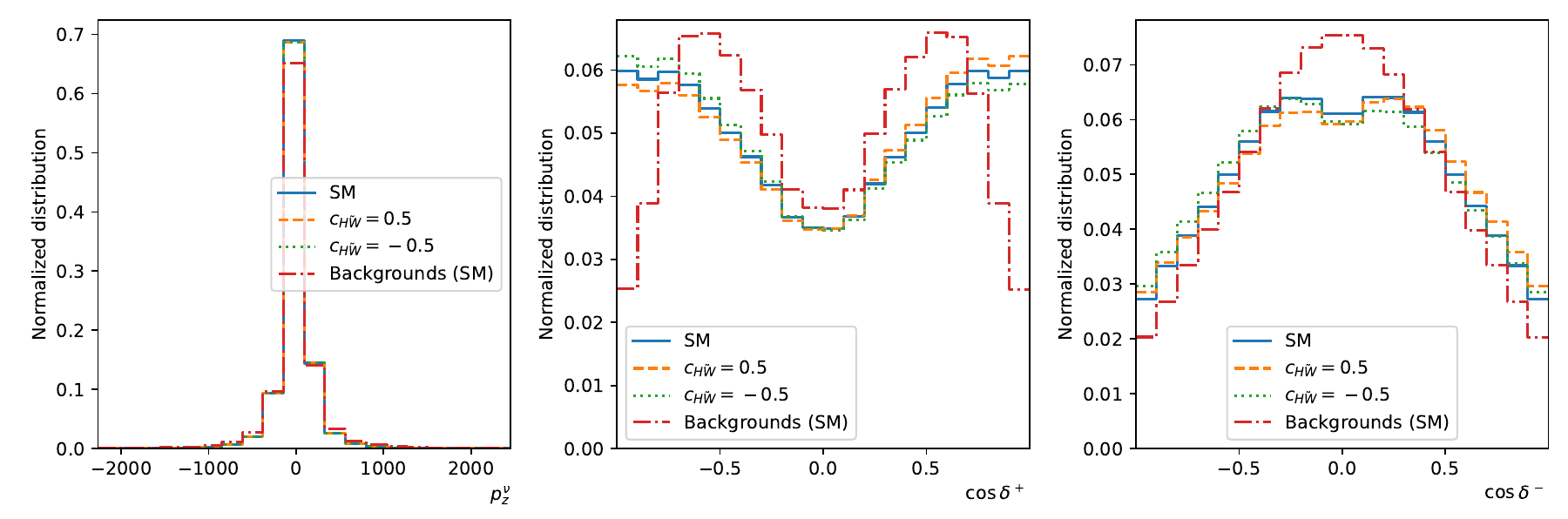}
	\caption{Distributions (normalized to unit area) of $\pznu$ (left), $\cDPlus$ (center) and $\cDMinus$ (right) for the SM backgrounds (red), SM signal (blue), and two signal samples with $\cHWtil=0.5$ (orange) and $\cHWtil=-0.5$ (green) obtained by morphing the benchmark signal samples.}
	\label{fig:pznuangularobservablesshapeonly}
\end{figure}

Several observables sensitive to the CP-odd interference component have been proposed in the literature \cite{GodboleHCP,MethodOfMoments}. In this work, we took as a benchmark the observables defined in Ref. \cite{GodboleHCP}, $\cDPlus$ and $\cDMinus$. These observables are defined in Eqs. \ref{eq:cosDPlus} and \ref{eq:cosDMinus}, respectively.
\begin{align}
		\cos \delta^+ & = \frac{\mathbf{p}_\ell^{(W)} \cdot \left( \mathbf{p}_H \times \mathbf{p}_W \right)}{|\mathbf{p}_\ell^{(W)}| |\mathbf{p}_H \times \mathbf{p}_W| } \label{eq:cosDPlus} \\
		\cos \delta^- &= \frac{\mathbf{p}_W \cdot (\mathbf{p}_\ell^{(-)} \times \mathbf{p}_\nu^{(-)} )}{|\mathbf{p}_W| |\mathbf{p}_\ell^{(-)} \times \mathbf{p}_\nu^{(-)}| }  \label{eq:cosDMinus} 
\end{align}
\noindent where $\mathbf{p}_W$ and $\mathbf{p}_H$ are the momenta of the W and Higgs bosons, respectively, $\mathbf{p}_\ell^{(W)}$ is the momenta of the charged lepton in the W boson rest frame, and $\mathbf{p}_{(\ell/\nu)}^{(-)}$ are the momenta of the charged lepton/ neutrino in the rest frame of the Higgs with $(\mathbf{p}_H \rightarrow - \mathbf{p}_H)$. To reconstruct the longitudinal momentum of the neutrino, $\pznu$, we identify $\ptnu \equiv \MET$, and solve the quadratic equation $p_{W_\mu} p_W^\mu = m_W^2$, neglecting imaginary parts which can arise from resolutions effects. Following the recommendations of Refs. \cite{GodboleHCP, BoostedVHbb2020ATLAS}, we choose the solution which minimizes the difference in longitudinal boosts between the W and of the Higgs boson, $|\beta_z^H - \beta_z^W|$, where $\beta_z = p_z / \sqrt{p_z^2 + m^2}$. We show in Fig. \ref{fig:pznuangularobservablesshapeonly} the distributions (normalized to unit area) of the longitudinal momentum of the neutrino and of the described angular observables, for the SM signal and backgrounds, as well as two signals with $\cHWtil \neq 0$.

\begin{figure}[h!]
	\centering
	\includegraphics[width=0.75\linewidth]{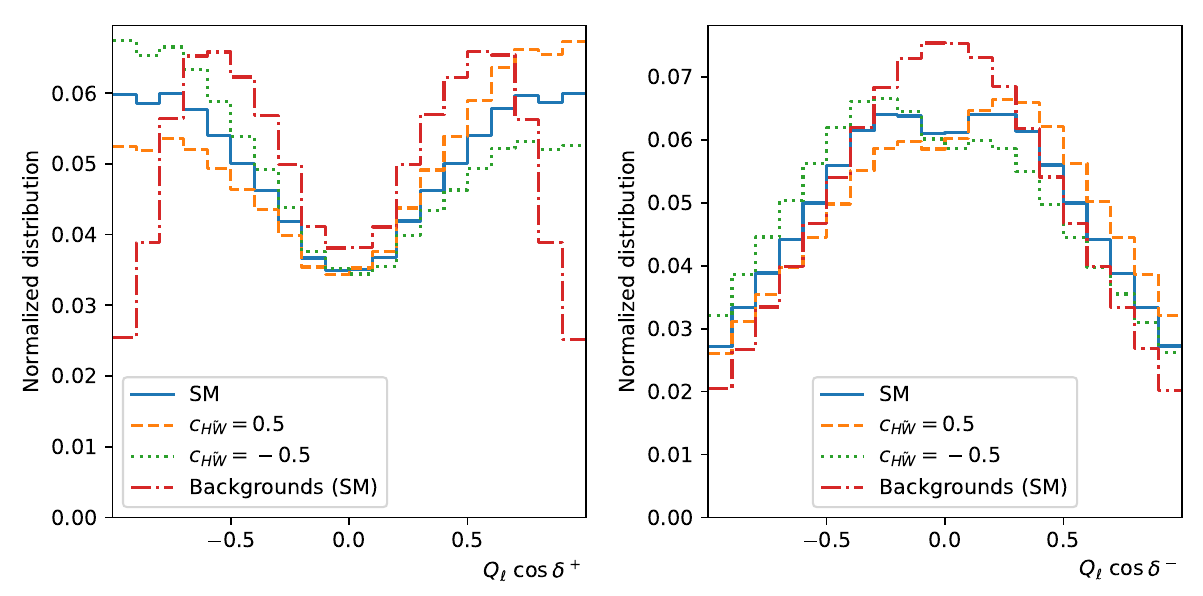}
	\caption{Distributions (normalized to unit area) of $\qlCDPlus$ (center) and $\qlCDMinus$ (right) for the SM backgrounds (red), SM signal (blue), and two signal samples with $\cHWtil=0.5$ (orange) and $\cHWtil=-0.5$ (green) obtained by morphing the benchmark signal samples.}
	\label{fig:angularobservables_chargeweighted_shapeonly}
\end{figure}

We observed that the asymmetry in the angular observable distributions for $\cHWtil \neq 0$ had the opposite sign for events with W bosons with opposite charges. This phenomenon happens because the charged lepton (and neutrino) in $W^+H$ and $W^-H$ have opposite helicities, leading to some of the angular components having an opposite sign, an idea we derived from Ref. \cite{Pre_MethodOfMoments_HZZ}. This effect leads to a dilution of the asymmetry in charge-inclusive distributions, which become more similar to those of the SM and backgrounds, reducing their sensitivity to CP-violating effects. To mitigate this effect, we propose a similar observable, obtained by multiplying the angular observable's value by the corresponding lepton charge, $Q_\ell$. In Fig. \ref{fig:angularobservables_chargeweighted_shapeonly}, we show the distributions of $\qlCDPlus$ and $\qlCDMinus$ (normalized to unit area) and observe that these maintain the properties of the distributions $\cDPlus$ and $\cDMinus$ but with increased asymmetry, when compared to the distributions in Fig. \ref{fig:pznuangularobservablesshapeonly}.

\subsection{SALLY, data augmentation and training settings}\label{sec:sally_obs}


In this work, we used the SALLY method to build a detector-level optimal observable with the SM point ($\theta \equiv \cHWtil = 0$) as reference. We created an unweighted training dataset with 11.5M events drawn from a combined SM signal+background sample, with a probability given by the value of the event generator weight, calculating the joint score for each. We used neural networks with a single hidden layer with 50 hidden units, scaling the input variables to have zero mean and unit variance. The loss function is minimized with the AMSGrad \cite{AMSGrad} optimizer for 50 epochs, with a learning rate that decays exponentially from $10^{-3}$ to $10^{-4}$ and a batch size of 128. The dataset was divided into a training and validation set, which have 75\% and 25\% of the events, respectively. We used early stopping based on the validation loss to avoid overfitting. Similarly to Ref. \cite{MadMinerWH}, we trained an ensemble of five networks to make the prediction more robust to different neural network random seeds. The network input variables are the following:

\begin{itemize}
	\item 4-vector of the two \bquarks and the charged lepton
	\item $\pt$, $\eta$, $\theta$ and $\phi$ of the two \bquarks
	\item $\pt$, $\eta$, $\theta$ and $\phi$ of the Higgs boson candidate
	\item $ m_{bb}$
	\item $\Delta \phi$ and $\DeltaR$ between the two \bquarks
	\item $\Delta \phi$ and $\DeltaR$ between the lepton and each of the \bquarks
	\item $x$ and $y$ components of \MET
	\item $|\MET|$
	\item $\pt$ and $\phi$ of the W boson candidate 
	\item $\Delta \phi$ between \MET and each of the \bquarks
	\item $\Delta \phi$ between \MET and the lepton
\end{itemize}

\begin{figure}[ht]
	\centering
	\includegraphics[width=0.5\linewidth]{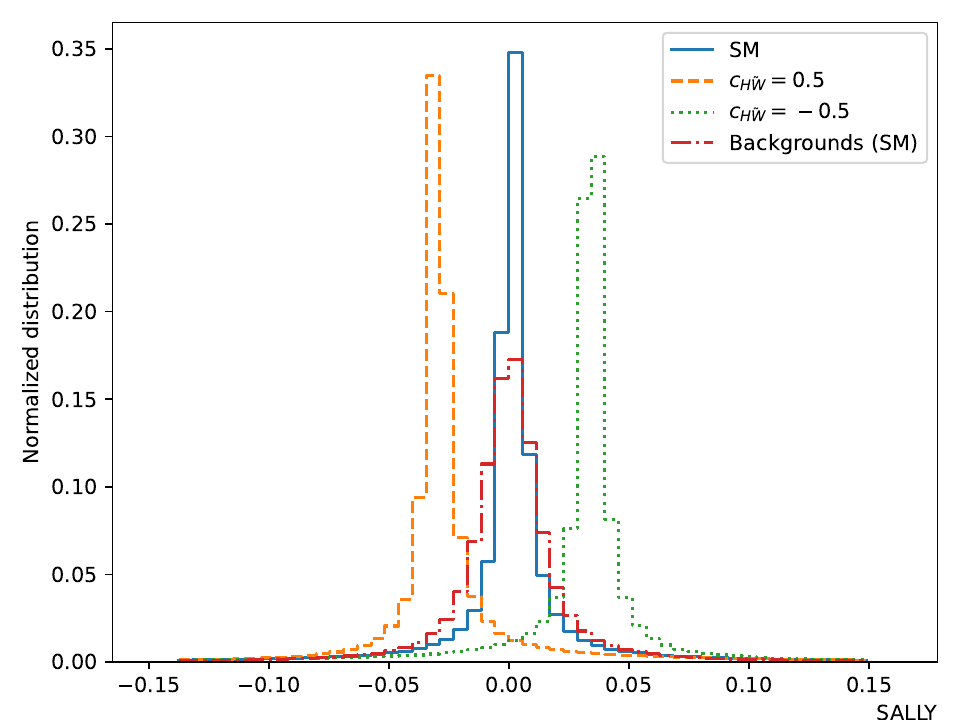}
	\caption{Distributions (normalized to unit area) of a SALLY observable trained on the signal+background sample for the SM backgrounds (red), SM signal (blue), and two signal samples with $\cHWtil=0.5$ (orange) and $\cHWtil=-0.5$ (green) obtained by morphing the benchmark signal samples.}
	\label{fig:sallykinematicvariablestraining}
\end{figure}

In Fig. \ref{fig:sallykinematicvariablestraining}, we show the SALLY observable distributions (normalized to unit area). We can see that we have distributions centered at 0 for SM signal and backgrounds and non-zero mean value for non-zero values of $\cHWtil$, with the sign of the asymmetry being opposite to the sign of the coefficient.

	\section{Statistical analysis}\label{sec:stat_analysis}

The likelihood ratio can be used in a frequentist setting to perform hypothesis testing and derive confidence intervals. One can expand the likelihood ratio around a reference point $\theta_{ref}$ as in Eq. \ref{eq:FisherInfo_in_LLR}.

\begin{equation}\label{eq:FisherInfo_in_LLR}
	-2 \mathbb{E} \left[\log \frac{ p_{\text{full}}(x|\theta)}{p_{\text{full}}(x|\theta_{ref})} \right] = - \mathbb{E} \left[ \frac{\partial^2 \log p_{\text{full}}(x|\theta)}{\partial \theta_i \partial \theta_j} \right]\rvert_{\theta_{ref}} \left(\theta - \theta_{ref}\right)_i \left(\theta - \theta_{ref}\right)_j + \mathcal{O}(\theta^3),
\end{equation}

\noindent where $p_{\text{full}}$ is the likelihood. 

$- \mathbb{E} \left[ \frac{\partial^2 \log p_{\text{full}}(x|\theta)}{\partial \theta_i \partial \theta_j} \right]\rvert_{\theta_{ref}} \equiv I_{ij}(\theta_{ref})$ (where $\mathbb{E}$ is the expected value notation) is the Fisher Information matrix, which quantifies the sensitivity of measurements around $\theta_{ref}$, with larger components indicating more precise measurements. The Fisher Information in histograms of observables is calculated from the variation of the differential cross-section in each bin (and will depend on the binning). If an estimator of the score (such as SALLY) is available, one can obtain the Fisher Information in the full kinematics from the estimated score for each event (regardless of binning).

Neglecting the $\mathcal{O}(\theta^3)$ terms leads to a likelihood ratio (and limits) linearized in $\theta$, and is called the Local Fisher Distance formalism \cite{InfoGeoHiggsMeasurements,InfoGeoHiggsCP,MadMinerWH}. To consistently take into account the higher order terms in the likelihood ratio, one can instead use the full likelihood ratio in the asymptotic formalism \cite{asymptotic_llr} or the Global Fisher Distance formalism \cite{MadMinerPhysicsComplete,MadMinerWH}. It is shown in Ref. \cite{InfoGeoHiggsMeasurements} that such limits are equivalent.

\section{Results}\label{sec:results}

To study the sensitivity of different observables, we compare the range of the expected 95\% CL limits on $\cHWtil$ (where a smaller the range is equivalent to a higher sensitivity), assuming the SM ($\cHWtil=0$) as the null hypothesis, for a luminosity of 300 $\ifb$, calculating the full likelihood ratio in steps of $\cHWtil=0.024$. We explore different binnings to study the sensitivity of these observables in different regions of phase space. We compare these limits with linearized limits derived using the Local Fisher Distance formalism to study the relative importance of the SM-EFT interference and quadratic EFT terms in the squared matrix element to the sensitivity of each observable. These are also compared to limits derived with the full likelihood ratio, but neglecting changes in the total rate. The results are shown in Table \ref{tab:resultsSTXS_SMonly_full_pythiadelphes}.

\begin{table}[h!]
		\centering
\setlength{\extrarowheight}{2mm}
	\resizebox{\linewidth}{!}{
	\begin{tabular}{lrrr}
		\textbf{Observable} & \textbf{Linearized limits} & \textbf{Full limits} & \makecell[r]{\textbf{Full limits}\\\textbf{(neglecting changes} \\ \textbf{in total rate)}}\\ \hline 
		$\ptw \in  \left[ 0.0 , 150.0 , 250.0 \right] \:\GeV$ & N.A. & $ [-0.336 , 0.336] $ &       $ [-0.360 , 0.336] $ \\
		$\ptw \in  \left[ 0.0 , 75.0 , 150.0 , 250.0 \right] \:\GeV$ &   N.A. & $ [-0.336 , 0.336] $ &       $ [-0.360 , 0.336] $ \\
		$\ptw \in  \left[ 0.0 , 75.0 , 150.0 , 250.0 , 400.0 \right] \:\GeV$ & N.A. & $ [-0.336 , 0.336] $ &      $ [-0.336 , 0.336] $ \\
		$\ptw \in  \left[ 0.0 , 75.0 , 150.0 , 250.0 , 400.0 , 600.0 \right] \:\GeV$ &   N.A. & $ [-0.336 , 0.336] $ &      $ [-0.336 , 0.336] $ \\[2mm]
		\makecell[l]{$\ptw \in  \left[ 0.0 , 75.0 , 150.0 , 250.0 , 400.0 \right] \:\GeV \otimes$ \\ $\mttot \in  \left[ 0.0 , 400.0 , 800.0 \right] \:\GeV$} & N.A. & $ [-0.312 , 0.312] $ &      $ [-0.336 , 0.336] $ \\[5mm]
		\makecell[l]{$\ptw \in  \left[ 0.0 , 75.0 , 150.0 , 250.0 , 400.0 , 600.0 \right] \:\GeV \otimes$ \\ $\mttot \in  \left[ 0.0 , 400.0 , 800.0 \right] \:\GeV$} &  N.A. & $ [-0.312 , 0.312] $ &      $ [-0.336 , 0.336] $ \\
		\hline
		$\qlCDPlus \in  \left[  -1.0 , 0.0 , 1.0\right]$ &   $[-0.356 , 0.356]$ &   $ [-0.360 , 0.360] $ &      $ [-0.408 , 0.408] $ \\
		$\qlCDPlus \in  \left[  -1.0 , -0.5 , 0.0 , 0.5  1.0\right]$ &   $[-0.321 , 0.321]$ & $ [-0.336 , 0.336] $ &        $ [-0.360 , 0.360] $ \\
		$\qlCDPlus \in  \left[  -1.0 , -2/3 , -1/3 , 0.0 , 1/3 , 2/3 , 1.0\right]$ &   $[-0.307 , 0.307]$ & $ [-0.312 , 0.312] $ &       $ [-0.336 , 0.360] $ \\[2mm]
		\makecell[l]{$\qlCDPlus \in  \left[  -1.0 , -2/3 , -1/3 , 0.0 , 1/3 , 2/3 , 1.0\right] \otimes$ \\ $\ptw \in  \left[ 0.0 , 75.0 , 150.0 , 250.0 , 400.0 , 600.0 \right] \:\GeV$} &   $[-0.137 , 0.137]$ & $ [-0.144 , 0.144] $ &      $ [-0.144 , 0.144] $ \\[5mm]
		\makecell[l]{$\qlCDPlus \in  \left[  -1.0 , -2/3 , -1/3 , 0.0 , 1/3 , 2/3 , 1.0\right] \otimes$ \\ $\mttot \in  \left[ 0.0 , 400.0 , 800.0 \right] \:\GeV$} &   $[-0.151 , 0.151]$ & $ [-0.168 , 0.168] $ &      $ [-0.168 , 0.168] $ \\\hline
		
		$\qlCDMinus \in  \left[  -1.0 , 0.0 , 1.0\right]$ &  $[-0.357 , 0.357]$ &   $ [-0.360 , 0.336] $ &      $ [-0.408 , 0.384] $ \\
		$\qlCDMinus \in  \left[  -1.0 , -0.5 , 0.0 , 0.5  1.0\right]$ &   $[-0.328 , 0.328]$ & $ [-0.336 , 0.336] $ &        $ [-0.360 , 0.360] $ \\
		$\qlCDMinus \in  \left[  -1.0 , -2/3 , -1/3 , 0.0 , 1/3 , 2/3 , 1.0\right]$ &   $[-0.317 , 0.317]$ & $ [-0.312 , 0.312] $ &       $ [-0.336 , 0.336] $ \\[2mm]
		\makecell[l]{$\qlCDMinus \in  \left[  -1.0 , -2/3 , -1/3 , 0.0 , 1/3 , 2/3 , 1.0\right] \otimes$ \\ $\ptw \in  \left[ 0.0 , 75.0 , 150.0 , 250.0 , 400.0 , 600.0 \right] \:\GeV$} &   $[-0.149 , 0.149]$ & $ [-0.192 , 0.120] $ &      $ [-0.192 , 0.120] $ \\[5mm]
		\makecell[l]{$\qlCDMinus \in  \left[  -1.0 , -2/3 , -1/3 , 0.0 , 1/3 , 2/3 , 1.0\right] \otimes$ \\ $\mttot \in  \left[ 0.0 , 400.0 , 800.0 \right] \:\GeV$} &   $[-0.160 , 0.160]$ & $ [-0.192 , 0.144] $ &      $ [-0.192 , 0.144] $ \\\hline		
		
		SALLY, w/ detector-level observables, 25 bins & $[-0.111, 0.111]$ & $[-0.192, 0.192]$ &  $[-0.168, 0.192]$ \\[3mm] 
		\makecell[l]{SALLY, w/ detector-level observables \\ $\in \left[  -1.0 , -2/3 , -1/3 , 0.0 , 1/3 , 2/3 , 1.0\right]$} & - & $[-0.120, 0.168]$ &  $[-0.120 , 0.168]$ \\[5mm] 
		\makecell[l]{SALLY, w/ detector-level observables + $\pznu$ and $\qlCDPlus$, \\ 25 bins} & $[-0.098, 0.098]$ & $[-0.192, 0.192]$ & $[-0.192 , 0.192]$ \\[3mm]
		\makecell[l]{SALLY, w/ detector-level observables + $\pznu$ and $\qlCDPlus$ \\ $\in  \left[  -1.0 , -2/3 , -1/3 , 0.0 , 1/3 , 2/3 , 1.0\right]$} & - & $[-0.168 , 0.192] $ & $[-0.168 , 0.192]$ \\ 
		\makecell[l]{SALLY, w/ detector-level observables + $\pznu$, $\qlCDPlus$, \\ $\qlCDMinus$, 25 bins} & $[-0.097, 0.097]$ & $[-0.192, 0.144]$ & $[-0.216,  0.144]$ \\[3mm]
		\makecell[l]{SALLY, w/ detector-level observables + $\pznu$, $\qlCDPlus$, \\ $\qlCDMinus$, $\in  \left[  -1.0 , -2/3 , -1/3 , 0.0 , 1/3 , 2/3 , 1.0\right]$} & - & $[-0.168,  0.192] $ & $[-0.168 , 0.192]$ \\ \hline		
	\end{tabular}
	}
	\caption{95\% CL on $\cHWtil$ assuming the SM $\cHWtil=0$ as the null hypothesis for the different observables and binnings tested (showing only lower bin edges, for the last bin the upper edge is $+\infty$) for an integrated luminosity of $\mathcal{L} = 300 \:\text{fb}^{-1}$.}
	\label{tab:resultsSTXS_SMonly_full_pythiadelphes}
\end{table}

In the upper section of Table \ref{tab:resultsSTXS_SMonly_full_pythiadelphes}, we show the limits derived from one-dimensional (1D) and two-dimensional (2D) histograms of two energy-dependent observables, $\ptw$ and $\mttot$. The linearized limits obtained with these observables are outside of the $\cHWtil$ range where the linearization of the likelihood ratio is a valid approximation (we write N.A. to make this clear). This strengthens the idea that these observables alone are sensitive to the quadratic EFT term in the likelihood ratio (and squared matrix element). Neglecting the cross-section information leads to negligible changes in the limits, showing that the change in the shape of the distributions is driving the sensitivity of such a measurement. Energy-dependent observables alone are unsuitable for probing CP-violating components since they cannot distinguish between the contribution from the quadratic component of a CP-odd operator and the contributions from other CP-even operators. 

In the middle section of Table \ref{tab:resultsSTXS_SMonly_full_pythiadelphes}, we show the limits derived from 1D histograms of the angular observables $\qlCDPlus$ and $\qlCDMinus$ as well as 2D histograms of an angular observable and one of the two energy-dependent observables, $\ptw$ and $\mttot$. The full limits obtained with an angular observable are looser than those obtained with $\ptw$ when using a coarse binning for the angular observable histogram. We explain this observation by the fact that $\ptw$ has a large sensitivity to the quadratic EFT term, which is negligible for the angular observables. 2D limits obtained with an energy-dependent variable and an angular observable are tighter than those obtained with an angular observable alone, both with the linearized and full likelihood ratio, an effect coming from the sensitivity of energy-dependent observables to the SM-EFT interference component when binning in a CP-odd angular observable, as shown in Ref. \cite{wh_lvaa_eft_fcc}, as well as additional sensitivity of energy-dependent observables to the quadratic term. The size of the linearized and full limits obtained with angular observables is very similar, indicating that most of their sensitivity comes from the linear component, which reinforces the point presented in Sec. \ref{sec:ang_obs} and in previous work. Finer binnings in the angular observable histograms lead only to a marginal increase in the sensitivity. The limits obtained with 2D histograms do not change when changes in the total cross-section are neglected, similar to histograms of energy-dependent observables.

On the bottom section of the table, we show the limits obtained with SALLY using the kinematic observable set described in Sec. \ref{sec:sally_obs}, as well as an additional models where we added $\pznu$, $\qlCDMinus$ and $\qlCDPlus$ as additional input variables. As mentioned in Sec. \ref{sec:stat_analysis}, we derive the linearized limits for SALLY using the estimated score for each event, so the binning mentioned pertains only to the histograms used to derive the full limits. We explore two different binnings: one with 25 bins between the 0.1\% and the 99.9\% percentile ($ \in[-0.79, 0.78]$) with equal fraction of (weighted) events in each bin (default definition in \texttt{MadMiner}), and another with 6 equally-spaced bins between -1 and 1. The full limits are about 70\% looser than the linearized ones, explained by SALLY being only optimal when quadratic EFT effects are subdominant. The limits obtained with SALLY are $\approx 40\%$ tighter than the ones obtained from a 1D histogram of an angular observable alone, showing that a multivariate SALLY method can capture very well kinematic correlations sensitive to CP-violating components, without having to reconstruct the neutrino longitudinal momenta. We obtained the tightest full limits for a 6-bin histogram of the SALLY model trained with standard kinematic observables as input variables.

%


	\section{Conclusions}\label{sec:conclusions}

This work studied CP-violating components in the HWW interaction in leptonic WH production, parametrized by an effective dimension-6 CP-odd operator, $\oHWtil$. We explored a machine-learning-based simulation-based inference method SALLY that estimates a detector-level optimal observable using the full kinematic information of the event. We compared the sensitivity of SALLY with that of 1D and 2D distributions of energy-dependent and angular observables. An analysis was put in place, including backgrounds as well as effects of parton shower, hadronisation and detector reconstruction. We showed that weighting angular observables by the charge of the lepton leads to an improvement in their sensitivity. To compare different observables, we extracted 95\% CL limits with the SM ($\cHWtil=0$) as the null hypothesis for a luminosity of 300 $\ifb$, both with linearized and full likelihood ratios. This allowed us to separate the effect of the SM-EFT interference and quadratic EFT terms on the sensitivity of the different observables.

We showed that a SALLY method trained around the SM point, with only detector-level observables as input has the best sensitivity to the linear component of all the observables studied, around 30\% higher than the sensitivity obtained from a 2D histogram of the transverse momentum of the W boson, $\ptw$ and the CP-odd angular observable $\qlCDPlus$. Taking into account the quadratic components consistently, the best sensitivity is obtained by a SALLY observable trained with standard kinematic observables as input variables in a 6-bin setup, slightly better than that of the 2D histogram of $\ptw$ and $\qlCDPlus$.

In conclusion, an observable such as SALLY can extract the maximal amount of information from the full event kinematics, leading to optimal limits.

Our analysis code is available online in Ref. \cite{sally_cpv_wh_code}. It is composed of a series of Python scripts using extensively the \texttt{MadMiner} Python package~\cite{MadMinerIntro,MadMiner_code}. More information and instructions on how to run the code can be found on the GitHub page. This code is made available not only for analysis preservation and replicability, but also so that any similar future work - such as exploring other processes or operators - can build on it simpler and faster.


\subsubsection*{Acknowledgments}
%
R.B. is supported by the Portuguese Foundation for Science and Technology (FCT) under Contracts no. CERN/FIS-PAR/0026/2021 and SFRH/BD/150792/2020 and would like to acknowledge Shankha Banerjee, Samuel Homiller, Stephen Jiggins and Ines Ochoa for all the insightful discussions during the developzment of this work. P.C.M. is partially supported by the Portuguese Foundation for Science and Technology (FCT) under Contracts no. CERN/FIS-PAR/0010/2021. R.S. is partially supported by the Portuguese Foundation for Science and Technology (FCT) under Contracts no. UIDB/00618/2020, UIDP/00618/2020, CERN/FIS-PAR/0025/2021, CERN/FIS-PAR/0010/2021 and CERN/FIS-PAR/0021/2021.

\vspace*{1cm}
\bibliographystyle{JHEP}
\bibliography{Bibliography}	
	

\end{document}